\documentclass{aa}  

\usepackage{graphicx}
\usepackage{txfonts}
\begin{document}

   \title{Detection of CI line emission from the detached CO shell of the AGB star R Sculptoris
   \thanks{Based on observations with the Atacama Pathfinder EXperiment (APEX) telescope. APEX is a collaboration between the Max-Planck-Institut f{\"u}r Radioastronomie, the European Southern Observatory, and the Onsala Space Observatory.}}
   \titlerunning{CI line emission towards R Scl}

   \author{H. Olofsson  \inst{1}
          \and          
           P. Bergman \inst{1}
           \and
          M. Lindqvist \inst{1}
          }

   \institute{Dept. of Earth and Space Sciences, Chalmers Univ. of Technology,
              Onsala Space Observatory, SE-43992 Onsala, Sweden\\
              \email{hans.olofsson@chalmers.se}   
             }

   \date{Received 15 June 2015; accepted 3 September 2015}

 \abstract{}{}{}{}{} 
 
  \abstract
   {Stars on the asymptotic giant branch (AGB) lose substantial amounts of matter, to the extent that they are important for the chemical evolution of, and dust production in, the universe. The mass loss is believed to increase gradually with age on the AGB, but it may also occur in the form of bursts, possibly related to the thermal pulsing phenomenon. Detached, geometrically thin, CO shells around carbon stars are good signposts of brief and intense mass ejection.}
   {We aim to put further constraints on the physical properties of detached CO shells around AGB stars.}
   {The photodissociation of CO and other carbon-bearing species in the shells leads to the possibility of detecting lines from neutral carbon. We have therefore searched for the CI($^3P_1-\,^3P_0$) line at 492\,GHz towards two carbon stars, S~Sct and R~Scl, with detached CO shells of different ages, $\approx$\,8000 and 2300 years, respectively.}
   {The CI($^3P_1-\,^3P_0$) line was detected towards R~Scl. The line intensity is dominated by emission from the detached shell. The detection is at a level consistent with the neutral carbon coming from the full photodissociation of all species except CO, and with only limited photoionisation of carbon. The best fit to the observed $^{12}$CO and $^{13}$CO line intensities, assuming a homogeneous shell, is obtained for a shell mass of $\approx$\,0.002\,$M_\odot$, a temperature of $\approx$\,100\,K, and a CO abundance with respect to H$_2$ of 10$^{-3}$. The estimated CI/CO abundance ratio is $\approx$\,0.3 for the best-fit model. However, a number of arguments point in the direction of a clumpy medium, and a viable interpretation of the data within such a context is provided.}
   {}


   \maketitle
%
\section{Introduction}

It is well established that low- and intermediate-mass stars go through extensive mass loss during the evolution on the asymptotic giant branch (AGB). Most of this mass loss probably occurs over an extended time scale through a wind that appears to increase in strength as the star evolves \citep{habi96}. However, there is also ample evidence of highly episodic mass loss. Most notable are the geometrically thin, detached CO and dust shells around carbon-rich AGB stars. Presently there are published data on seven CO shells \citep{schoetal05} and 17 dust shells of this type \citep[][and references therein]{ramsetal11, coxetal12, maeretal14, mecietal14} (not all of the dust shells are clearly of the geometrically thin type). We note here that some stars with prominent dust shells, e.g. AX~Cyg and RT~Cap \citep{kersetal10, mecietal14}, do not have detectable CO shells, most likely an effect of CO photodissociation if the shell is old enough. Even though most of the mass is not lost during these episodes, they provide an interesting phenomenon which may shed light on the mass-loss mechanism of AGB stars, e.g. the mass ejection may be connected to a physical process with well-defined variations of stellar characteristics (see below), the line of sight confusion is limited due to the thinness of the shells and this allows studies of the small-scale structure of the circumstellar medium, and the evolution of the gaseous and dusty components can be compared in some detail.

The reason for the episodic mass loss is not definitely identified. \citet{olofetal90, olofetal93a} provided good arguments, based on the detection statistics of an essentially complete sample of carbon stars, that the strong modulation of the mass-loss rate is driven by the changes that a star goes through during a He-shell flash, or thermal pulse. \citet{schoetal05} provided evidence that the interaction of a previous, slower stellar wind has an effect on the detached shell characteristics. The mass-loss modulation is estimated to be, at least, one order of magnitude, being less than 10$^{-6}$\,$M_\odot$\,yr$^{-1}$ before and after the shell ejection and larger than 10$^{-5}$\,$M_\odot$\,yr$^{-1}$ during shell ejection. The shell ejection time scale is a few hundred years \citep{olofetal00, maeretal12}. Curiously, no M-stars, which also go through thermal pulsing, have been found with detached CO and/or dust shells of this type despite several searches \citep{nymaetal92, kersolof99, kersetal10}.

The physical properties of the detached shells have been probed so far using CO radio line emission and through dust in either scattered stellar light or thermal emission. The CO lines have the advantage of providing information on the kinematics, and they can be studied at high angular resolution. Data of the latter type show remarkably spherical shells with width/radius ratios of <\,0.1 \citep{lindetal96, olofetal00, maeretal12}. The images in scattered light are insensitive to any excitation processes \citep{gonzetal01, olofetal10}, and in polarisation mode they provide important geometrical information \citep{gonzetal03, maeretal10, maeretal14, ramsetal11}. The images in thermal emission are slightly more sensitive to excitation (the grain temperature) and there is more line of sight confusion, but they provide good estimates of the dust mass if observed at longer wavelengths. It turns out that a determination of the physical characteristics (density, temperature, clumpiness, etc.) of a detached shell is difficult using the existing data. 

Hence, there is a need for additional probes, but in the case of molecular lines they are hard to find owing to efficient photodissociation. Eventually, CO is also dissociated into C and O, though it is much more stable than other species owing to line dissociation and self-shielding \citep{mamoetal88}. Therefore, the CI($^3P_1 -\, ^3P_0$) line at 492 GHz is a possible alternative probe of the shell characteristics. It is, however, well known that this line is not easily detected in AGB and post-AGB objects. There are published detections in only one carbon star (CW~Leo, \citet{keenetal93}), one supergiant ($\alpha$~Ori, \citet{huggetal94}), and seven carbon-rich post-AGB objects [CRL618, CRL2688, HD56126, HD44179 (Red Rectangle), NGC6720 (Ring nebula), NGC7027, NGC7293 (Helix nebula), \citet{bachetal94, youn97, younetal97, knapetal00}].

In this paper we report on a search for the CI($^3P_1 -\, ^3P_0$) line in two detached CO shell sources, R~Scl and S~Sct, where the shell ages differ by about a factor of four from the former ($\approx$\,2300 yr) to the latter ($\approx$\,8000 yr) \citep{schoetal05}. In the case of R~Scl the search was successful, and we discuss the implications of the CI detection for our understanding of detached CO shells around AGB stars.

\section{Observations}

We have used the APEX 12\,m telescope located at an altitude of about 5100\,m  in the Chilean Andes \citep{gustetal06} to observe the CI($^3P_1 -\, ^3P_0$) line at 492 GHz. The Swedish heterodyne facility instrument APEX-3 \citep{vassetal08}, a double sideband receiver, was used together with the facility FFT spectrometer covering about 4\,GHz. The observations were made in April and June 2014 under programme O-093.F-9306.

The observed sources are presented in Table~\ref{t:sources} with variable type (V), period ($P$), stellar luminosity ($L_{\star}$), and distance ($D$) for each star. The periods, luminosities, and distances, which are estimated by fitting the spectral energy distributions and calculating the luminosities from the period-luminosity relations, are taken from \citet{gonzetal01} and \citet{schoetal05}.

The observations were made in a dual beamswitch mode where the source is alternately placed in the signal and the reference beam, using a beam throw of about 2$\arcmin$. Regular pointing checks were made on strong CO line emitters (e.g. in the case of R~Scl on the object itself) and continuum sources. Typically, the pointing was found to be consistent with the pointing model within $\approx$\,3$\arcsec$. 

The raw spectra are stored in $T_{\mathrm A}^{\star}$-scale and converted to main-beam brightness temperature using $T_{\mathrm{mb}}$\,=\,$T_{\mathrm A}^{*}/\eta_{\mathrm{mb}}$. The antenna temperature, $T_{\mathrm A}^{\star}$, is corrected for atmospheric attenuation using a calibration unit, and $\eta_{\mathrm{mb}}$ is the main-beam efficiency. The adopted beam efficiency and the FWHM of the main beam at the observing frequency of 492~GHz are 0.6 and 13$\arcsec$, respectively. The uncertainty in the absolute intensity scale is estimated to be about $\pm 20$\%. The data were reduced by removing a first-order polynomial baseline using XS\footnote{XS is a package developed by P. Bergman to reduce and analyse single-dish spectra. It is publicly available from {\tt ftp://yggdrasil.oso.chalmers.se}}.

We note here that frequency shifts of the two isotope $^{13}$CI($^3P_1 -\, ^3P_0$) hyperfine lines with respect to the frequency of the $^{12}$CI($^3P_1 -\, ^3P_0$) line are small, 0.5\,MHz and 3.6\,MHz for the $F$\,=\,1/2--1/2 and 3/2--1/2 components, respectively. This means that in the circumstellar case where the lines are broad, emission from the $^{13}$C lines will contribute to the flux measured for the $^{12}$C line. However, since the circumstellar $^{12}$CO/$^{13}$CO is about 45 and 20 for S~Sct and R~Scl, respectively, we expect this contribution to be small \citep{bergetal93, vlemetal13}.

\begin{table}
\caption{Source information: variable type (V), period ($P$), luminosity ($L_{\star}$), and distance ($D$). }  
\label{t:sources}      
\centering                 
\begin{tabular}{l c c c  c}       
\hline\hline                
Source         & V   & $P$  & $L_{\star}$    & $D$ \\    
	           &     & [days]  & [L$_{\odot}$]  & [pc] \\
\hline 

\object{R~Scl} & SRb & 370  & 4900           & 370 \\                 
\object{S~Sct} & SRb & 148  & 4300           & 400 \\

\hline                                   
\end{tabular}
\end{table}

\section{Observational results}

\begin{figure}
\centering
   \includegraphics[width=6cm]{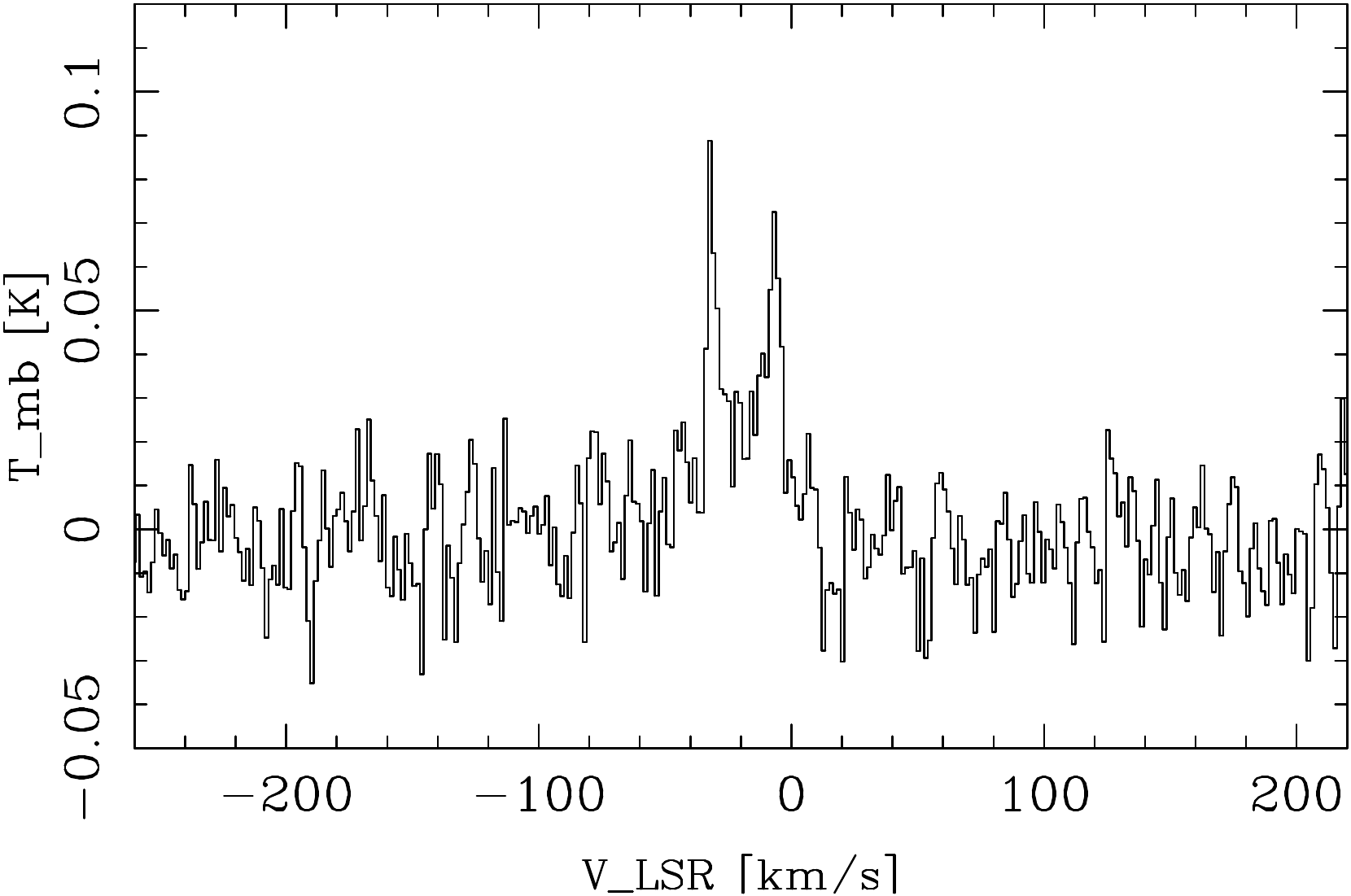}\\
   \includegraphics[width=6cm]{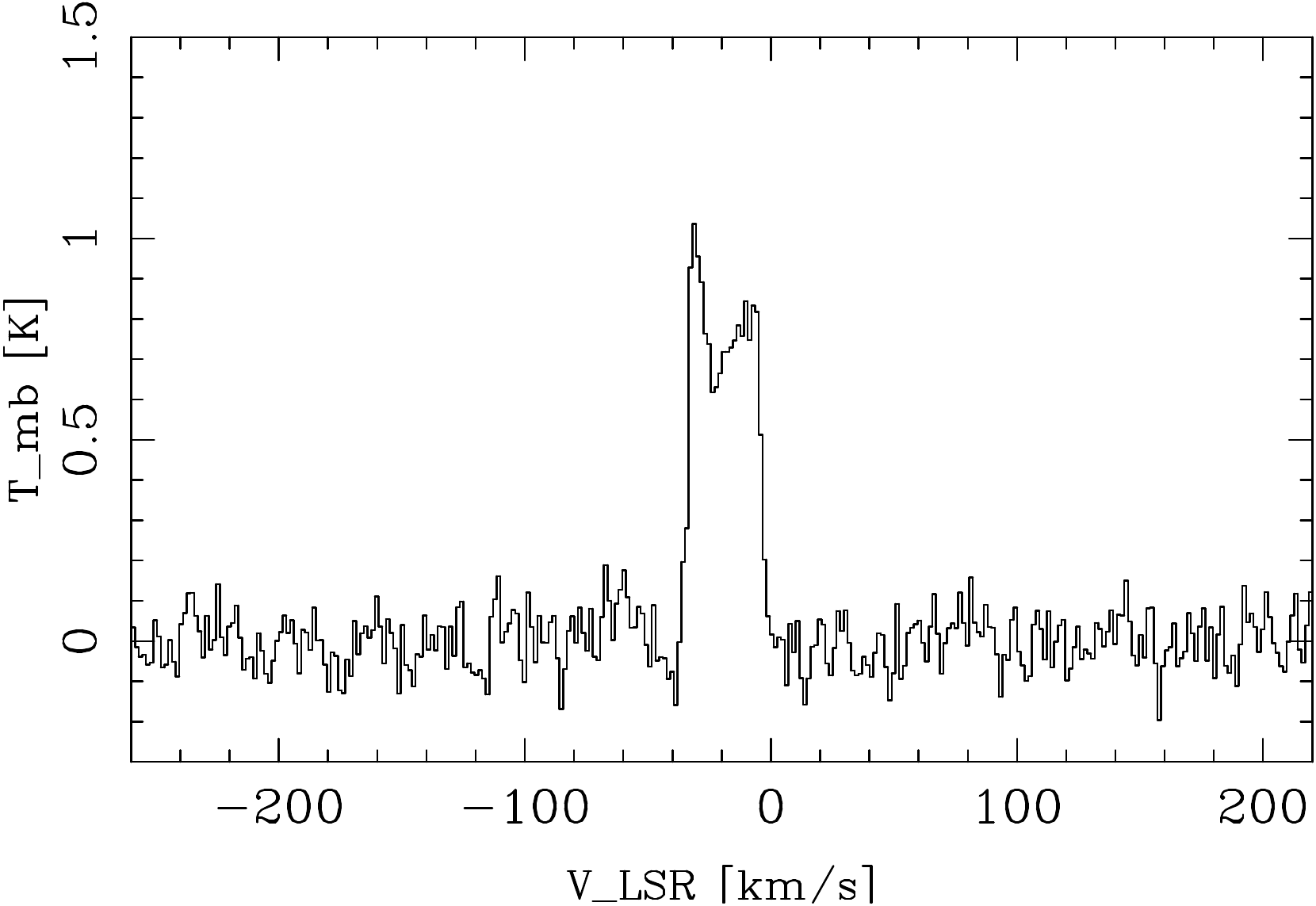}
      \caption{Observed CI($^3P_1 -\, ^3P_0$) (top) and CO($J$\,=\,4--3) (bottom) lines towards R Scl.
              }
     \label{f:spectra}
\end{figure}

The observed CI($^3P_1 -\, ^3P_0$) line towards R~Scl, binned to a velocity resolution of 1.5\,km\,s$^{-1}$, is presented in Fig.~\ref{f:spectra} together with the CO($J$\,=\,4--3) line at 461 GHz, which is observed with roughly the same beam width. The line profile has been fitted to the following function,
\begin{equation}
T(\upsilon) = T(\upsilon_{\rm sys})\,\left[ 1 - \left( \frac{\upsilon - \upsilon_{\rm sys}}{\upsilon_{\rm e}} \right)^2\right]^\alpha
\end{equation}
where $\upsilon_{\rm sys}$ is the central velocity, and $\upsilon_{\rm e}$ the gas expansion velocity of the CI line emission. A negative $\alpha$ means a double-peaked profile. The line intensities are integrated over the line profile. In the case of S~Sct, a 3$\sigma$ upper limit to the line intensity is estimated. A summary of the results are given in Table~\ref{t:obsresults}. The systemic and expansion velocities we obtain for CI agree well with those obtained from the CO lines in the case of R~Scl \citep{olofetal96}. Both lines are double-peaked, but the CI line has a markedly higher peak/centre ratio than the CO(4--3) line ($\alpha$\,=\,--1.1 compared to --0.4), which indicates that this line is significantly more dominated by emission from the detached shell (see Sect.~\ref{s:cI}).

\begin{table}
\caption{CI($^3P_1 -\, ^3P_0$) observational results}  
\label{t:obsresults}      
\centering                 
\begin{tabular}{l c c c c c}       
\hline\hline                
Source & $\upsilon_{\rm sys}$ & $\upsilon_{\rm e}$ & $\alpha$ & $T_{\rm peak}$ & $I$ \\    
	   &    [km\,s$^{-1}$]    & [km\,s$^{-1}$]       &          &  [K]           & [K\,km\,s$^{-1}$] \\
\hline 

\object{R~Scl} &    --19.0    &       14.5           & --1.1    &    0.09        &  1.5 \\                 
\object{S~Sct} &              &                      &          &                & <\,0.2 \\

\hline                                   
\end{tabular}
\end{table}

In addition to these data, we have assembled CO line data from the literature to constrain our modelling of the detached shell characteristics. These constitute the $^{12}$CO SEST $J$\,=\,(\mbox{1--0}) line \citep{olofetal96}, IRAM (\mbox{1--0}) and (\mbox{2--1}) lines \citep{danietal15}, APEX (\mbox{2--1}) and (\mbox{3--2}) lines \citep{debeetal10}, JCMT (\mbox{3--2}) and (\mbox{4--3}) lines \citep{schoetal05} and Herschel/HIFI (\mbox{5--4}) and (\mbox{9--8}) lines \citep{danietal15} towards R~Scl, and the SEST $J$\,=\,(\mbox{1--0}), (\mbox{2--1}) and (\mbox{3--2}) lines \citep{olofetal96} and APEX (\mbox{3--2}) line \citep{debeetal10} towards S~Sct. In addition, as further constraints on the R~Scl modelling (primarily on the optical depths) we have included the $^{13}$CO IRAM $J$\,=\,(\mbox{1--0}) line \citep{danietal15}, SEST (\mbox{1--0}) and (\mbox{2--1}) lines \citep{olofetal96} and APEX (\mbox{3--2}) line \citep{vlemetal13}.

\section{Radiative transfer}

We have used a radiative transfer code based on the accelerated lambda iteration method \citep[see e.g.][]{maeretal08} to derive quantitative results. As the geometrical (radius, $R_{\rm s}$, and width, $\Delta R_{\rm s}$) and kinematical (systemic velocity, $\upsilon_{\rm sys}$, and expansion velocity, $\upsilon_{\rm e}$) properties of the detached shells we adopt the results obtained by \citet{schoetal05} and \citet{maeretal12}. These are summarised in Table~\ref{t:shells}. The detached shells are assumed to be homogenous in density and kinetic temperature. The turbulent velocity width, $\upsilon_{\rm t}$, is chosen to be 1 and 0.8\,km\,s$^{-1}$ for R~Scl and S~Sct, respectively.

\begin{table}
\caption{Detached shell: gas parameters}  
\label{t:shells}      
\centering                 
\begin{tabular}{l c c c c c}       
\hline\hline                
Source      &   $\upsilon_{\rm sys}$   & $\upsilon_{\rm e}$   & $\upsilon_{\rm t}$  & $R_{\rm s}$           & $\Delta R_{\rm s}$   \\    
	        &    [km\,s$^{-1}$]        & [km\,s$^{-1}$]       & [km\,s$^{-1}$]      & [cm]                  & [cm]   \\
\hline 

\object{R~Scl} &  --18.9  &       14.3            &   1.0               & 1$\times$10$^{17}$  &   7$\times$10$^{15}$ \\                 
\object{S~Sct} &   $\phantom{1}$15.0  &       16.5            &   0.8               & 4$\times$10$^{17}$  &   1$\times$10$^{16}$ \\

\hline                                   
\end{tabular}
\end{table}

In the $^{12}$CO and $^{13}$CO excitation analysis, we included radiative transitions between rotational energy levels up to $J$\,=\,14 for the vibrational ground state (radiative excitation by a central source is negligible in the detached shells). The energy levels, transition frequencies, and Einstein $A$ coefficients were taken from the Cologne database for molecular spectroscopy [CDMS, \citet{mulletal05}]. Collisional rates were taken from \citet{yangetal10}, who calculated them separately for CO collisions with para- and ortho-H$_2$. The rates were weighed together assuming an ortho-/para-H$_2$ ratio of 3. They cover temperatures from 2 to 3000\,K. 

Data for CO assembled from the literature are used to constrain the circumstellar models. In the case of R~Scl we have ten and four lines of $^{12}$CO and $^{13}$CO, respectively, in the radiative transfer analysis (Table~\ref{t:coinputrscl}), and in the case of S~Sct we have four $^{12}$CO lines (Table~\ref{t:coinputssct}). 

\begin{figure*}
\centering
   \includegraphics[width=8cm]{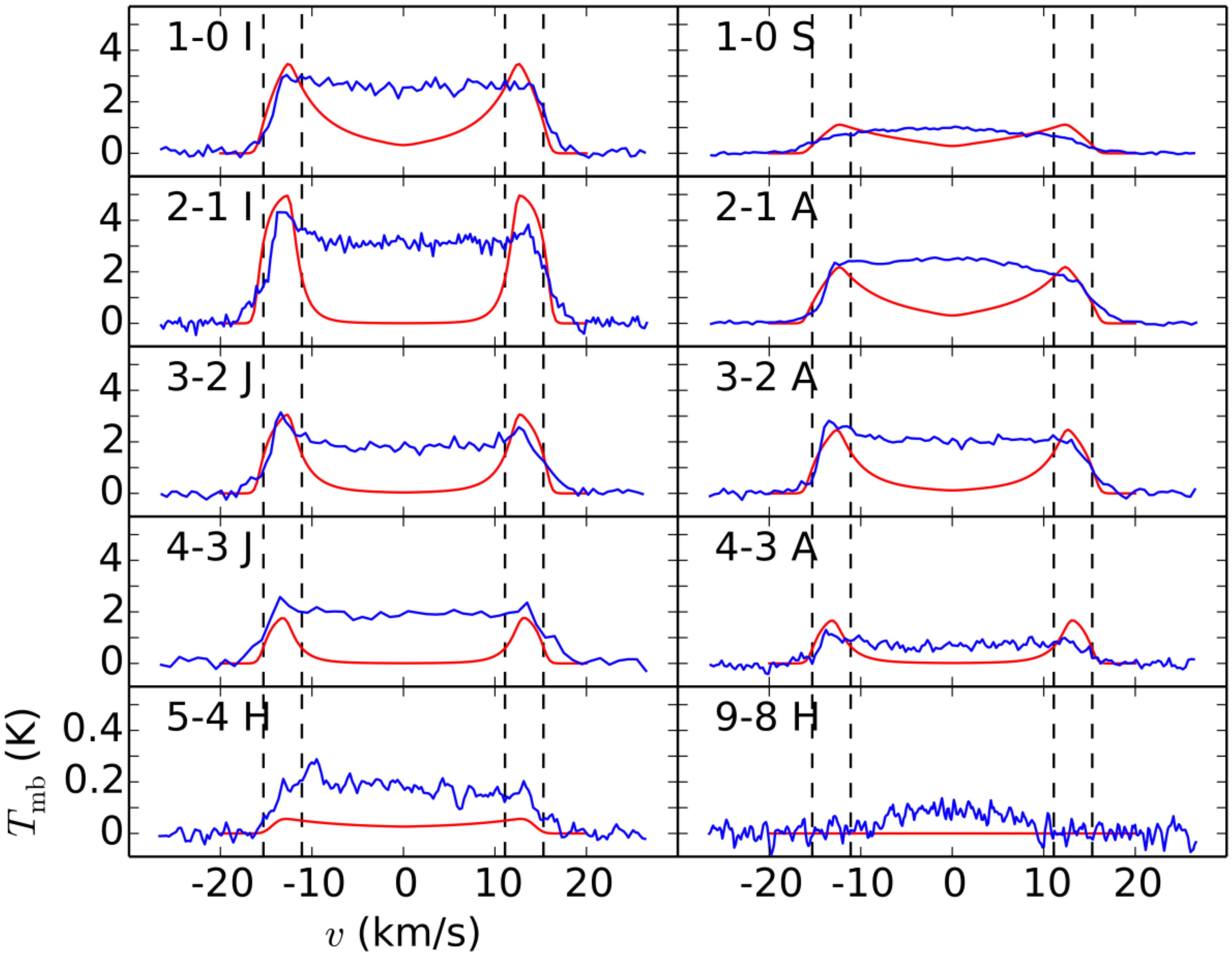} \hspace{5mm} \includegraphics[width=8cm]{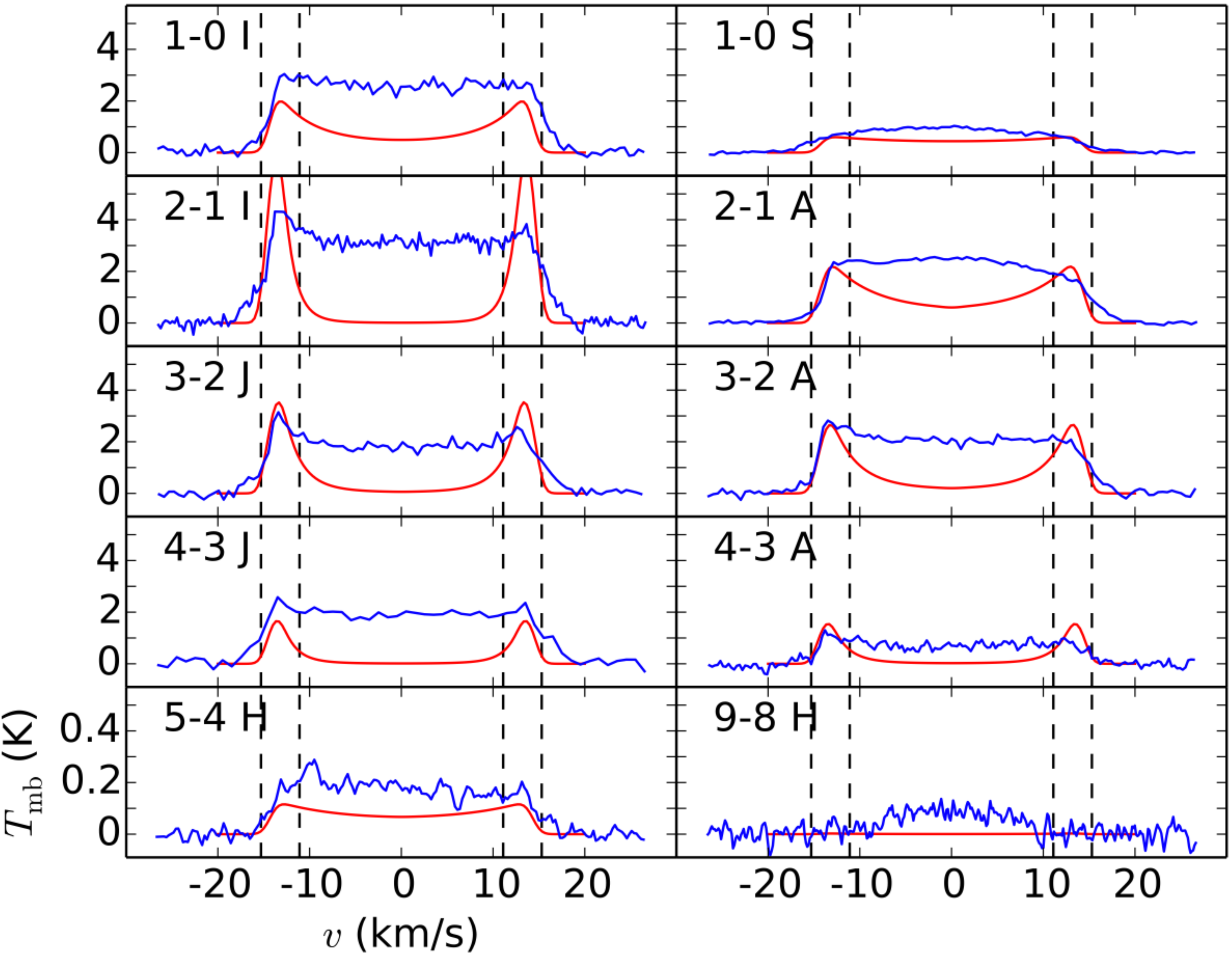}
      \caption{Observed (blue) and model (red) $^{12}$CO lines towards R~Scl (A=APEX, H=HIFI, I=IRAM, J=JCMT, S=SEST). The left panel corresponds to a 3$\times$10$^{5}$\,cm$^3$/10\,K model, and the right panel to a 2$\times$10$^{3}$\,cm$^3$/100\,K model. In both cases $f_{\rm CO}$\,=\,10$^{-3}$. The velocity scale is centred on the systemic velocity estimated to be \mbox{--18.9}\,km\,s$^{-1}$. The dashed lines indicate the regions used in the fitting to the line intensities.
              }
     \label{f:cosperscl}
\end{figure*}

The CO line emission of detached-shell sources consists of at least two components, the emission from the present-day mass loss and the emission from the detached shell \citep{olofetal96}. These components are normally well separated in velocity space since the velocity of the present-day wind is normally low ($\approx$\,5\,km\,s$^{-1}$), i.e. about 30\,\% of the detached shell expansion velocity \citep{olofetal96}, e.g. in the case of S~Sct. However, in the case of R~Scl the situation is more complicated. The present-day wind velocity is $\approx$\,60\% of the detached shell expansion velocity (a good estimate of the former is obtained from the $^{12}$CO($J$\,=\,9--8) line profile, $\approx$\,8.5\,km\,s$^{-1}$, see Fig.~\ref{f:cosperscl}). In addition, as shown by \citet{maeretal12} using ALMA  $^{12}$CO($J$\,=\,\mbox{3--2}) data, there is emission in the region between the present-day mass-loss envelope and the detached shell (in the form of a spiral pattern), and in the form of weak wings at velocities up to $\approx$\,19\,km\,s$^{-1}$ with respect to the systemic velocity. We have therefore chosen to perform a fit only to the integrated (over velocity) line intensities in the ranges 0.75\,$\upsilon_{\rm e}$\,$\le$\,$\mid$\,$\upsilon$-$\upsilon_{\rm sys}$\,$\mid$\,$\le$\,1.05\,$\upsilon_{\rm e}$. The integrated intensities in these ranges are listed in Table~\ref{t:coinputrscl}. The CO input data for S~Sct are given in Table~\ref{t:coinputssct}.

In the CI excitation analysis, we included the three ground-state levels $^3P_0$, $^3P_1$, and $^3P_2$. The collisional data were taken from \citet{schretal91}.

\begin{table}
\caption{Integrated CO line intensities for R Scl in the velocity ranges 11.1\,$\le$\,$\mid$\,$\upsilon$-$\upsilon_{\rm sys}$\,$\mid$\,$\le$\,15.3\,km\,s$^{-1}$}  
\label{t:coinputrscl}      
\centering                 
\begin{tabular}{l c c c c}       
\hline\hline                
Line         & Tel.  & Beam      &  $I_{\rm obs}$\tablefootmark{a}        & ref.\tablefootmark{b} \\    
	         &       & [\arcsec ] &  [K\,km\,s$^{-1}$]       & \\
\hline 

$^{12}$CO($J$\,=\,1--0) & IRAM                   &  21  &    21.6\,$\pm $\,0.3       &  1) \\
$^{12}$CO($J$\,=\,1--0) & SEST                   &  45  &    $\phantom{2}$4.6\,$\pm $\,0.1        &  2) \\
$^{12}$CO($J$\,=\,2--1) & APEX                   &  27  &    28.2\,$\pm $\,0.2       &  3) \\                
$^{12}$CO($J$\,=\,2--1) & IRAM                   &  11  &    14.2\,$\pm $\,0.1       &  1) \\
$^{12}$CO($J$\,=\,3--2) & APEX                   &  18  &    17.9\,$\pm $\,0.2       &  3) \\
$^{12}$CO($J$\,=\,3--2) & JCMT                   &  14  &    16.1\,$\pm $\,0.2       &  4) \\
$^{12}$CO($J$\,=\,4--3) & APEX                   &  14  &    16.4\,$\pm $\,0.5       &  5) \\
$^{12}$CO($J$\,=\,4--3) & JCMT                   &  12  &    $\phantom{2}$6.7\,$\pm $\,0.2        &  4) \\
$^{12}$CO($J$\,=\,5--4) & HIFI\tablefootmark{c}  &  36  &    $\phantom{2}$1.2\,$\pm $\,0.02       &  1) \\
$^{12}$CO($J$\,=\,9--8) & HIFI\tablefootmark{c}  &  20  &    $\phantom{2}$0.09\,$\pm $\,0.04    &  1) \\
$^{13}$CO($J$\,=\,1--0) & IRAM                   &  22  &    $\phantom{2}$2.0\,$\pm $\,0.09          &  1) \\
$^{13}$CO($J$\,=\,1--0) & SEST                   &  47  &    $\phantom{2}$0.45\,$\pm $\,0.02          &  2) \\
$^{13}$CO($J$\,=\,2--1) & SEST                   &  23  &    $\phantom{2}$1.3\,$\pm $\,0.04          &  2) \\
$^{13}$CO($J$\,=\,3--2) & APEX                   &  19  &    $\phantom{2}$1.9\,$\pm $\,0.05          &  6) \\ 

\hline                                   
\end{tabular}
\tablefoot{
\tablefoottext{a}{The errors are 1$\sigma$ noise limits.}
\tablefoottext{b}{1) \citet{danietal15}, 2) \citet{olofetal96}, 3) \citet{debeetal10}, 4) \citet{schoetal05}, 5) this paper, 6) \citet{vlemetal13}}
\tablefoottext{c}{The HIFI instrument on board the Herschel Space Observatory.}
}
\end{table}

\begin{table}
\caption{Integrated CO line intensities for SSct in the velocity ranges 11\,$\le$\,$\mid$\,$\upsilon$-$\upsilon_{\rm sys}$\,$\mid$\,$\le$\,19\,km\,s$^{-1}$}  
\label{t:coinputssct}      
\centering                 
\begin{tabular}{l c c c c}       
\hline\hline                
Line         & Tel.  & Beam      &  $I_{\rm obs}$\tablefootmark{a}      & ref.\tablefootmark{b} \\    
	         &       & [\arcsec ] &  [K\,km\,s$^{-1}$]   \\
\hline 
                 
$^{12}$CO($J$\,=\,1--0) & SEST  &  45       &    4.4\,$\pm $\,0.06                 &  1 \\
$^{12}$CO($J$\,=\,2--1) & SEST  &  23       &    3.0\,$\pm $\,0.07                 &  1 \\
$^{12}$CO($J$\,=\,3--2) & APEX  &  18       &    0.56\,$\pm $\,0.1                 &  2 \\
$^{12}$CO($J$\,=\,3--2) & SEST  &  15       &    1.0\,$\pm $\,0.09                 &  1 \\

\hline                                   
\end{tabular}
\tablefoot{
\tablefoottext{a}{The errors are 1$\sigma$ noise limits.}
\tablefoottext{b}{1) \citet{olofetal96}, 2) \citet{debeetal10}}
}
\end{table}

\begin{figure*}
\centering
   \includegraphics[width=5.85cm]{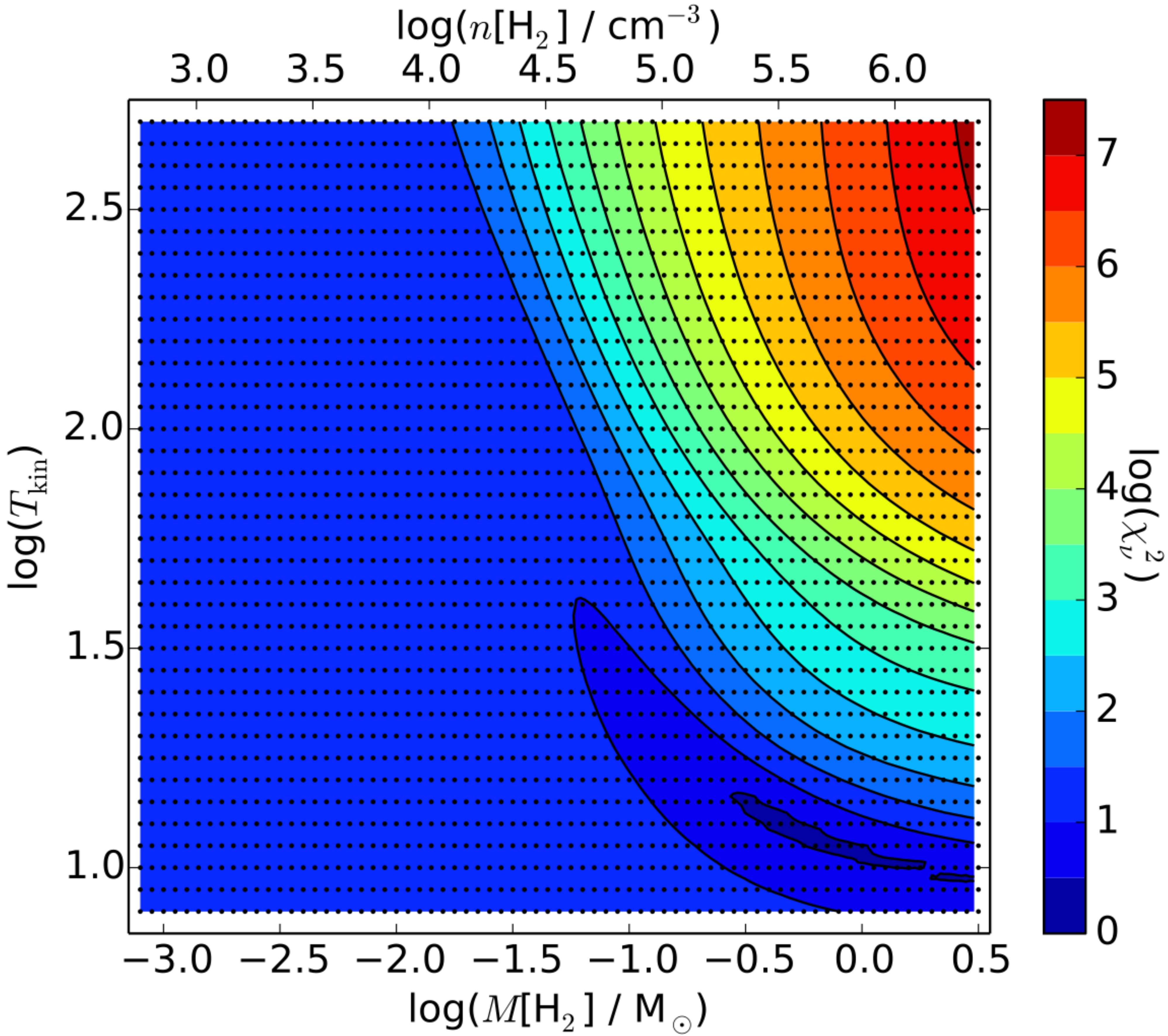} \includegraphics[width=6cm]{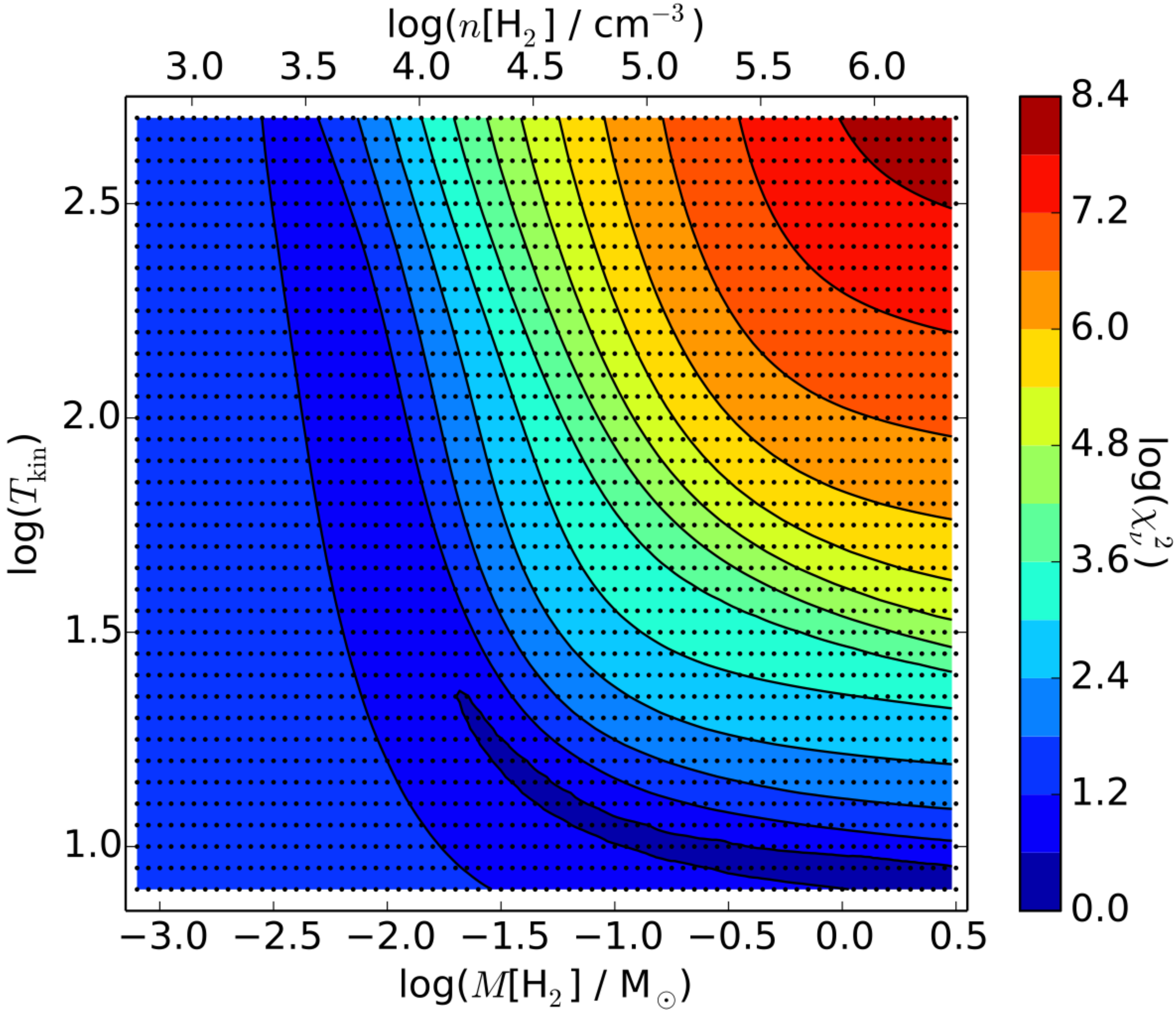} 
   \includegraphics[width=6cm]{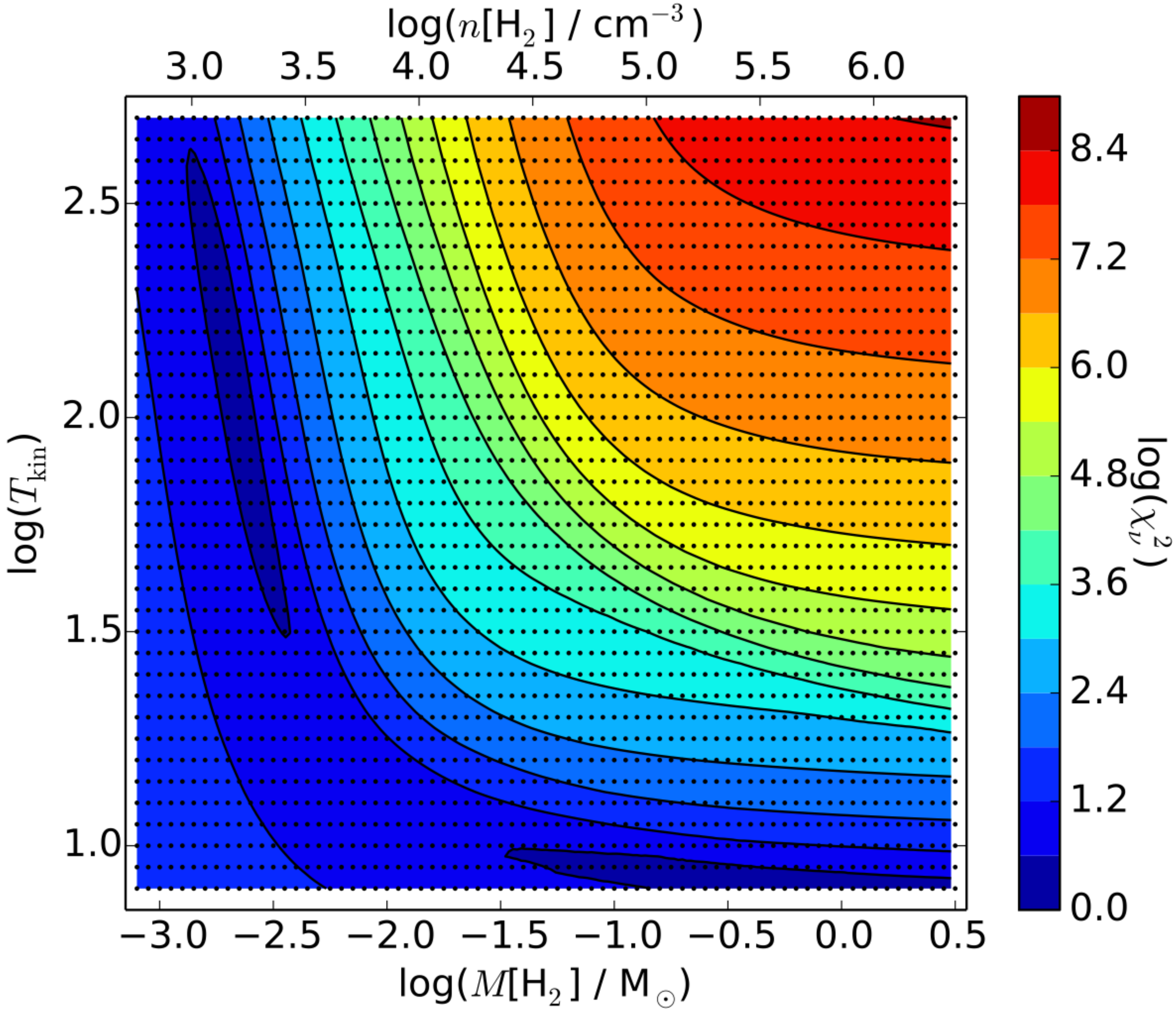}
      \caption{$\chi_{\rm red}^2$ maps (in logarithmic scale) for different $^{12}$CO abundances [10$^{-5}$ (left), 10$^{-4}$ (middle), 10$^{-3}$ (right)], where the density and kinetic temperature of the detached shell are free parameters, in the case of R~Scl.
              }
     \label{f:chi2rscl}
\end{figure*}

The best-fit models were found using a $\chi^2$ statistic defined as
\begin{equation}
\chi^2 = \sum_{1}^N \frac{(I_\mathrm{mod} - I_\mathrm{obs})^2}{\sigma^2}
\end{equation}
where $I$ is the integrated line intensities of the emission of the detached shell as observed (defined in Tables~\ref{t:coinputrscl} and \ref{t:coinputssct}) and in the models, $N$ the number of observed lines, and $\sigma$ the uncertainty of the intensity measurements, generally assumed to be 20\,\% unless otherwise noted. The reduced $\chi^2$ value is given by $\chi^2_\mathrm{red} = \chi^2/(N-p)$, where $p$ is the number of free parameters. In the determination of the detached-shell characteristics using the $^{12}$CO line data, the free parameters are the gas density and kinetic temperature. The $^{12}$CO abundance (with respect to H$_2$), $f_{\rm CO}$, is also unknown, so we provide results for a number of models with different values for this parameter. In the CI line emission modelling of the single observed line, the circumstellar model is fixed by the $^{12}$CO line modelling, and the only remaining free parameter is the CI abundance with respect to H$_2$, $f_{\rm CI}$. Similarly, for the $^{13}$CO line modelling we use the $^{12}$CO model results and vary the $^{13}$CO abundance, $f_{\rm {13}CO}$, until a best fit is obtained for a particular $^{12}$CO model.

\section{Results}

\subsection{Mass, density, and kinetic temperature of the detached shell}

In the case of R~Scl, we have run models for five different CO abundances, $f_{\rm CO}$\,=\,(0.01, 0.03, 0.1, 0.3, 1)$\times$10$^{-3}$. The last abundance applies to a carbon-rich medium where all the oxygen is locked up in CO (assuming a solar oxygen abundance of 5$\times$10$^{-4}$ \citep{grev09}), while the first applies to a medium where for example photodissociation has lead to a considerably lowered CO abundance. The $\chi_{\rm red}^2$ results are shown in Fig.~\ref{f:chi2rscl}. As can be clearly seen, reasonable fits (considering the uncertainties in extracting only the detached-shell emission) are found in large regions of the ($n_{\rm H_2}$,$T_{\rm k}$)-plane. The solutions with the lowest $\chi_{\rm red}^2$ values are found along a curve from high-densities/low-temperatures to low-densities/high-temperatures (its position depends on the CO abundance), and acceptable solutions are found over broad ranges in density and temperature to the left of this curve. The high-density/low-temperature solutions correspond to optically thick line emission, and detached-shell masses that are much higher than acceptable (see Sect.~\ref{s:mass}). For higher CO abundances, solutions of low-density/high-temperature models also provide good fits to the data. Results for some representative models are given in Table~\ref{t:modresults}. The predicted line profiles for one low-density/high-temperature model and one high-density/low-temperature model (both with $f_{\rm CO}$\,=\,10$^{-3}$) are shown in Fig.~\ref{f:cosperscl}. 

\begin{figure}
\centering
   \includegraphics[width=8cm]{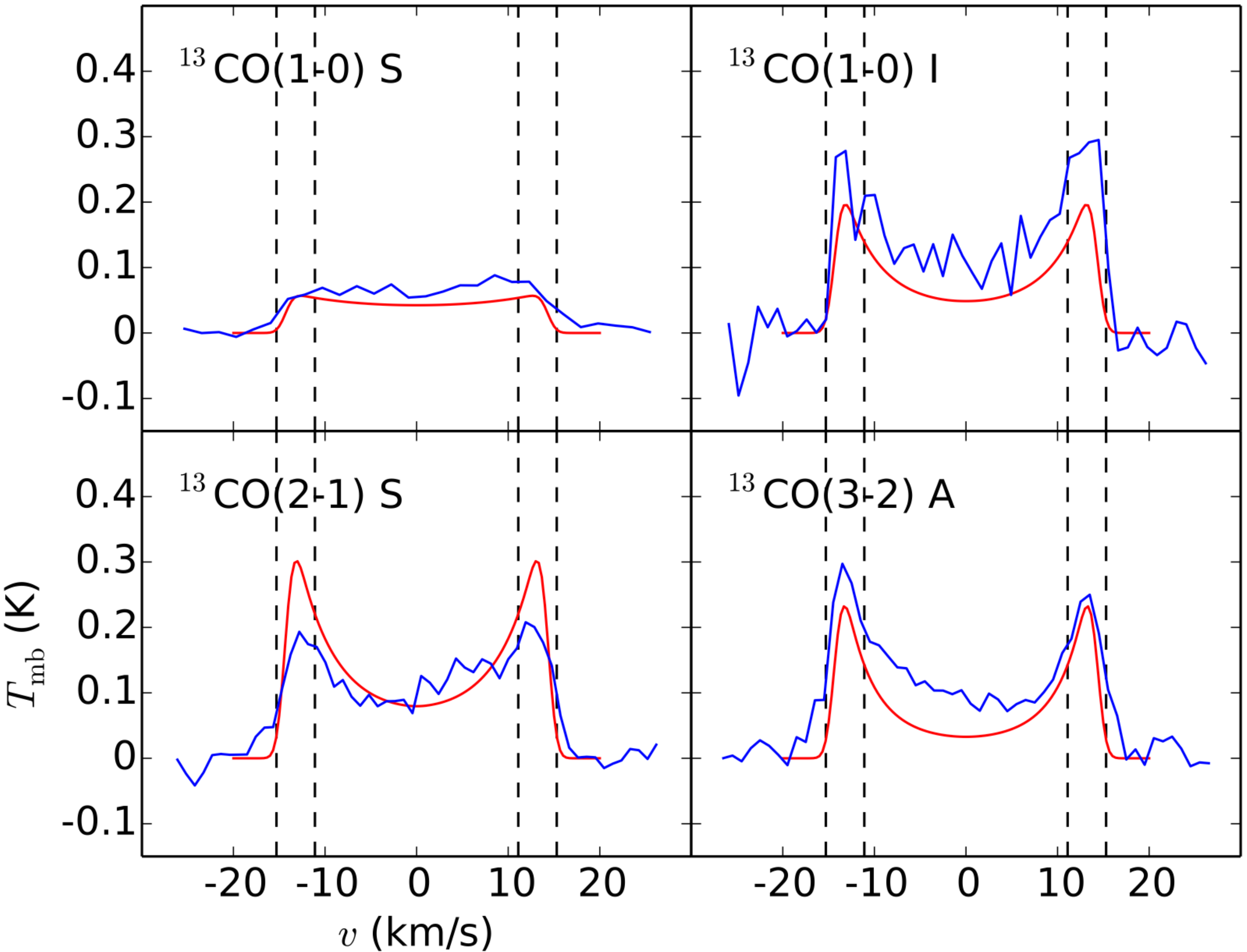}
      \caption{Observed (blue) and model (red) $^{13}$CO lines towards R~Scl (A=APEX, I=IRAM, S=SEST). The 2$\times$10$^{2}$\,cm$^3$/100\,K  and $f_{\rm CO}$\,=\,10$^{-3}$ model together with a $^{12}$CO/$^{13}$CO ratio of 13 is used in calculating the model spectra. The velocity scale is centred on the systemic velocity estimated to be \mbox{--18.9}\,km\,s$^{-1}$. The dashed lines indicate the regions used in the fitting to the line intensities.
              }
     \label{f:13cosperscl}
\end{figure}

We have also determined the best-fit $^{12}$CO/$^{13}$CO ratios for the presented models. The resulting  ratios are given in Table~\ref{t:modresults}, and the predicted $^{13}$CO line profiles for one low-density/high-temperature model with $f_{\rm CO}$\,=\,10$^{-3}$ is shown in Fig.~\ref{f:13cosperscl}. It is clear that the high-density/low-temperature models result in CO isotopologue ratios that are unlikely. This is discussed further in Sect.~\ref{s:mass}.

One can conclude from this that with only the CO lines as constraints, and within the adopted homogeneous-shell model, it is difficult to estimate independently the density and kinetic temperature in the detached shell. The acceptable parameter space is large, but the low-density/high-temperature solutions, only present in the high CO abundance case, are the preferred option.

\begin{table*}
\caption{Results for some representative models}  
\label{t:modresults}      
\centering                 
\begin{tabular}{l c c c c c c c}       
\hline\hline                
Source   & $f_{\rm CO}$       & $n_{\rm H_2}$      & $M_{\rm H_2}$   & $T_{\rm kin}$  & $\chi_{\rm red}^2$ & $^{12}$CO/$^{13}$CO  & CI/CO \\    
	     &                    & [cm$^{-3}$]        & [$M_\odot$]     & [K]      \\
\hline 

R Scl    & 1$\times$10$^{-5}$ & $\phantom{1.}$2$\times$10$^{6}$  & 3$\phantom{.0000}$    &  $\phantom{0}$10   & 3.2     & 120  & 0.056 \\ 
         & 1$\times$10$^{-4}$ & $\phantom{1.}$7$\times$10$^{5}$  & 1$\phantom{.0000}$    &  $\phantom{0}$10   & 6.3     & 390  & 0.017 \\                 
         & 1$\times$10$^{-3}$ & $\phantom{1.}$3$\times$10$^{5}$  & 0.04$\phantom{00}$    &  $\phantom{0}$10   & 3.9     & 150              & 0.045 \\
         & 1$\times$10$^{-3}$ & 2.4$\times$10$^{3}$              & 0.0035               &  $\phantom{0}$35   & 3.7     & $\phantom{0}$17  & 0.22$\phantom{0}$ \\
         & 1$\times$10$^{-3}$ & 1.5$\times$10$^{3}$              & 0.0022               & 100                & 2.4     &  $\phantom{0}$12  & 0.35$\phantom{0}$ \\
         \\         
S Sct    & 1$\times$10$^{-4}$ & $\phantom{1.}$2$\times$10$^{3}$  & 0.07$\phantom{00}$    &  $\phantom{0}$20   & 1.6     &      & <\,0.7 \\ 
         & 1$\times$10$^{-3}$ & $\phantom{1.}$2$\times$10$^{2}$  & 0.007$\phantom{0}$     & 100                & 1.8     &      & <\,1.1\\         

\hline                                   
\end{tabular}
\end{table*} 

\begin{figure*}
\centering
   \includegraphics[width=6cm]{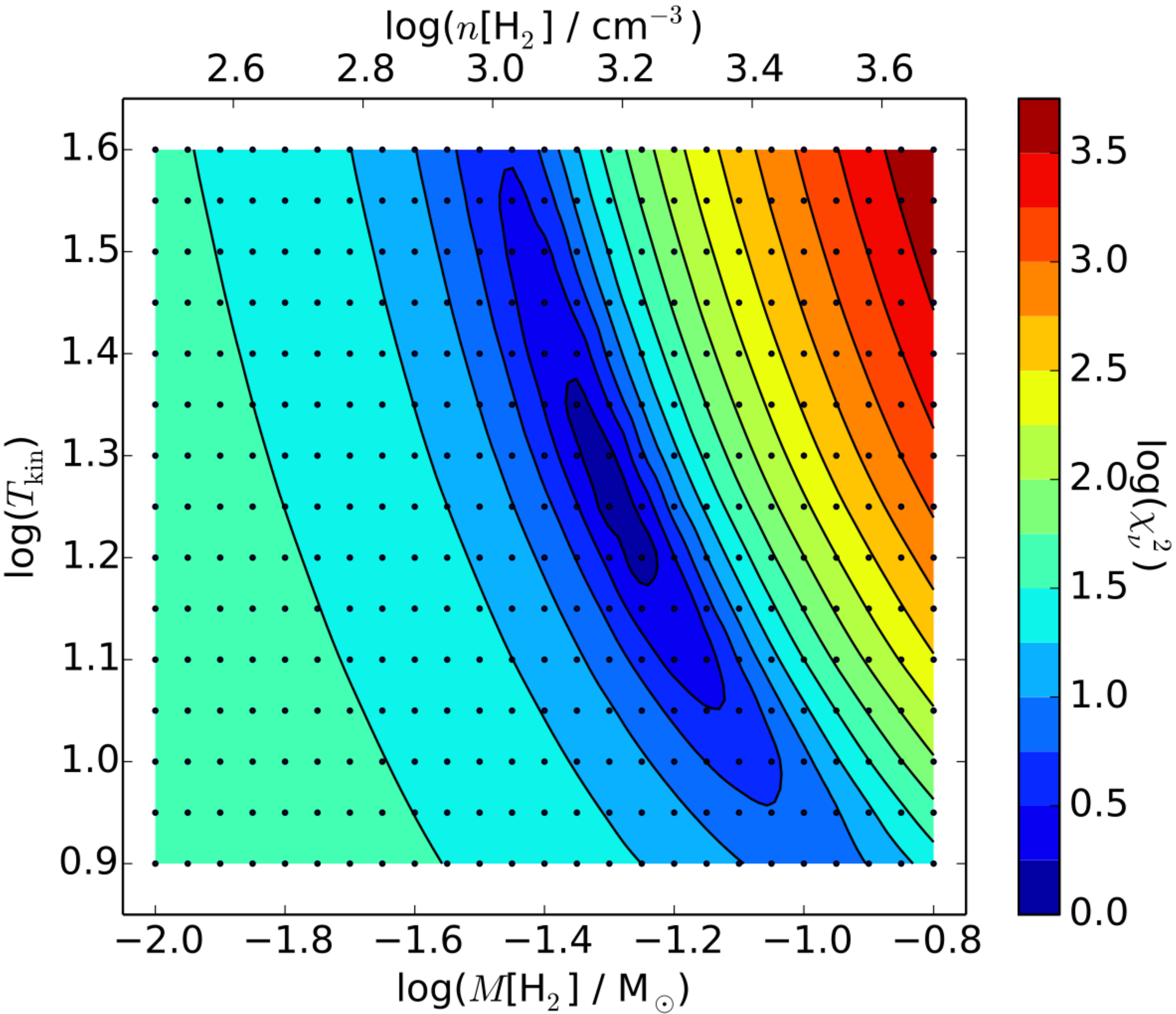} \includegraphics[width=6cm]{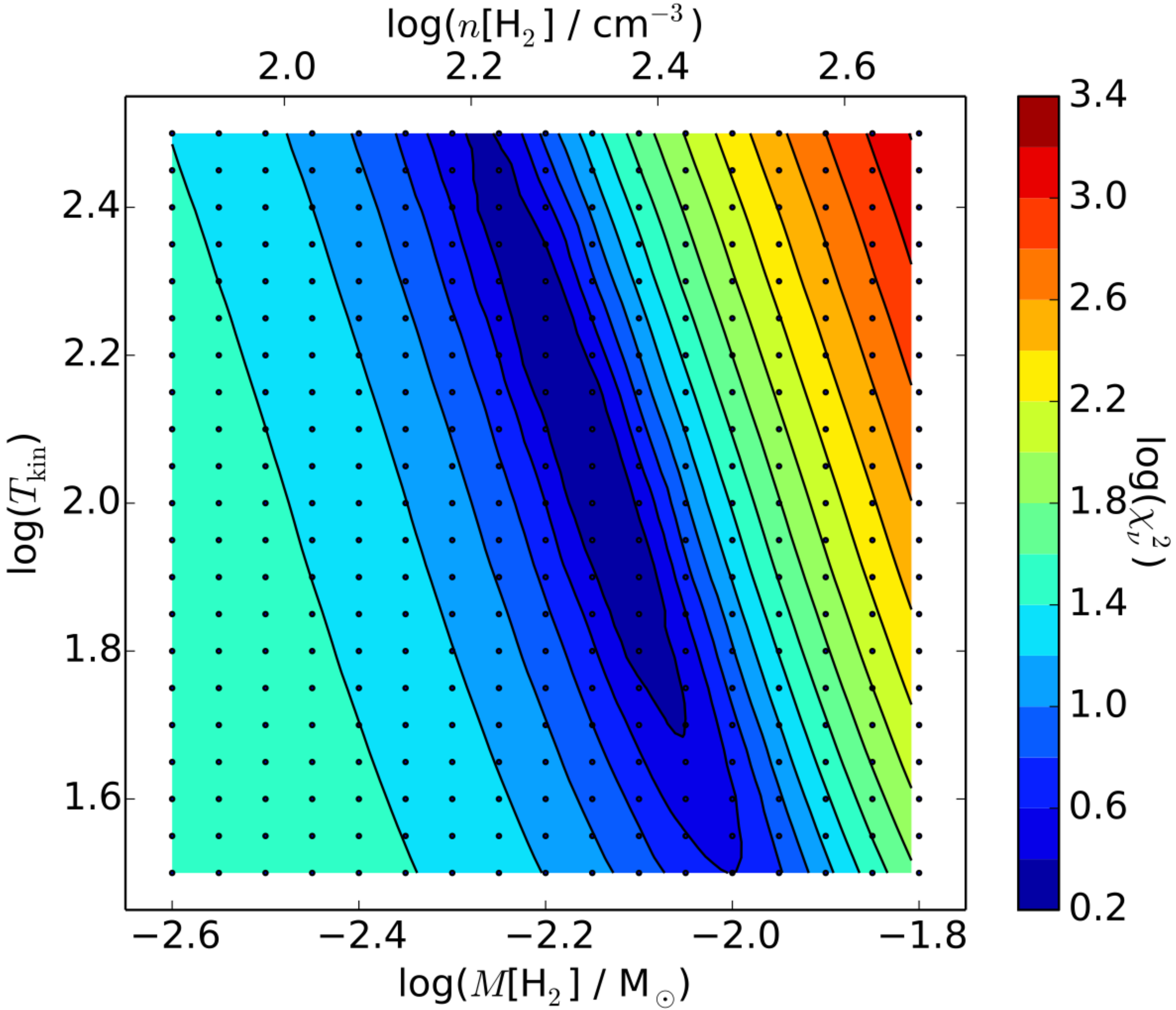} 
      \caption{$\chi_{\rm red}^2$ maps (in logarithmic scale) for different CO abundances [10$^{-4}$ (left), 10$^{-3}$ (right)], where the density and kinetic temperatures of the detached shell are free parameters, in the case of S~Sct.
              }
     \label{f:chi2ssct}
\end{figure*}

We have performed the same analysis for S~Sct based on the $^{12}$CO input data given in Table~\ref{t:coinputssct}. We restrict ourselves to only two CO abundances in this case, 10$^{-4}$ and 10$^{-3}$. The results are shown in Fig.~\ref{f:chi2ssct}. The general characteristics of the results are the same as those for R~Scl, but the parameter space is more confined in this case, and the $\chi_{\rm red}^2$ of the best-fit models are lower by about a factor of two. The latter is most likely an effect of the difficulty in isolating only the emission from the detached shell in the case of R~Scl.

\subsection{CI/CO abundance ratio}
\label{s:cI}

\begin{figure}
\centering
   \includegraphics[width=8cm]{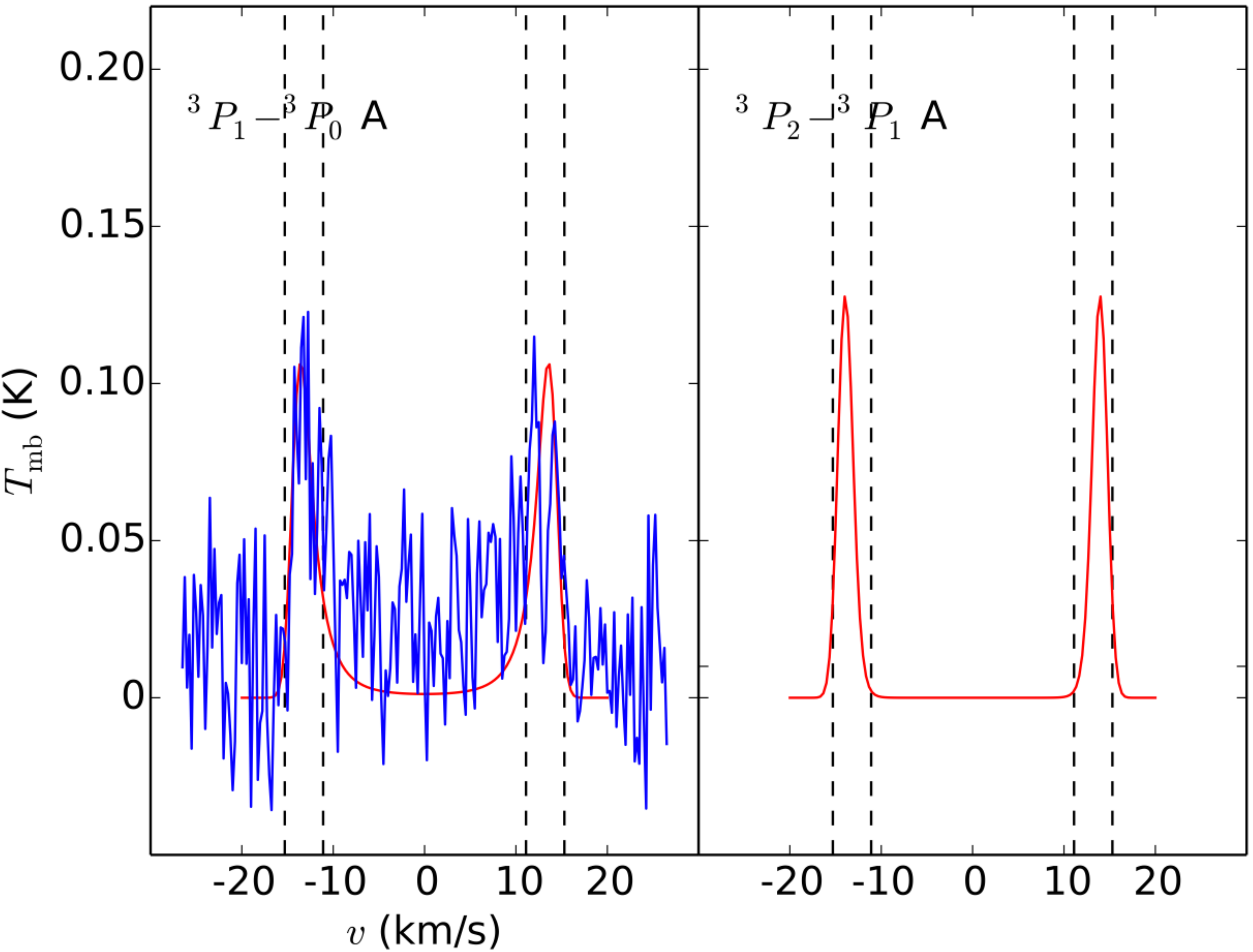}
      \caption{{\bf Left:} The CI($^3P_1 -\, ^3P_0$) observed (blue) and modelled (red) line spectrum. {\bf Right:} The CI($^3P_2 -\, ^3P_1$) modelled line spectrum. The 2$\times$10$^{2}$\,cm$^3$/100\,K  and $f_{\rm CO}$\,=\,10$^{-3}$ model is used, together with $f_{\rm CI}/f_{\rm CO}$\,=\,0.35, to calculate the model spectra. The velocity scale is centred on the systemic velocity estimated to be --18.9\,km\,s$^{-1}$. The dashed lines indicate the regions used in the fitting to the line intensities.
              }
     \label{f:cIsperscl}
\end{figure}

Based on the R Scl circumstellar model, constrained by the $^{12}$CO line data, we have estimated the abundance of neutral C compared to that of H$_2$ in its detached shell. It turns out that within a rather broad range of parameters (CO abundance, density, and kinetic temperature) the CI/CO abundance ratio is relatively constant, and lies in the range 0.2\,--\,0.5 for the models with low to moderate optical depths in the $^{12}$CO lines, i.e. the lower-density/higher-temperature models in Table~\ref{t:modresults}. The resulting CI($^3P_1 -\, ^3P_0$) and the predicted CI($^3P_2 -\, ^3P_1$) line profiles are shown in Fig.~\ref{f:cIsperscl} for the 2$\times$10$^3$\,cm$^{-3}$/100\,K/$f_{\rm CO}$\,=\,10$^{-3}$ model. The fit to the observed CI($^3P_1 -\, ^3P_0$ spectrum is good, indicating that most of the CI line emission comes from the detached shell, but it is noteworthy that a not insignificant amount of CI($^3P_1 -\, ^3P_0$) emission may also come from the present-day slower wind.

In the case of S~Sct we can only derive upper limits to the CI/CO abundance ratio, which are of the order of $\la$\,1 (see Table~\ref{t:modresults}). This is not a particularly constraining limit, but it indicates that a substantial amount of the CO molecules in the detached shell are still not photodissociated unless the majority of the carbon atoms have been ionised.

\section{Discussion}

As has been shown, we can fit the observational CO data within a broad range of densities (and hence detached-shell masses) and temperatures. However, there are a number of arguments, presented below in the case of R~Scl, that will guide us towards the most reasonable description of the detached shell.

\subsection{Kinetic temperature of the detached shell}

The kinetic temperatures in ``normal'' circumstellar envelopes have been well constrained through the modelling of CO radio lines \citep{olof03}. Using these results, we estimate the kinetic temperatures at a distance of 10$^{17}$\,cm from the star to be of the order 10\,K. Thus, even though the temperatures in the detached shells may be elevated above this value, for example due to interaction with a previous wind, it appears unlikely that they are much warmer than about 100\,K, which is also the preferred radiative transfer solution in the density/temperature space.

\subsection{Mass and line optical depths of the detached shell}
\label{s:mass}

The detached-shell masses become very high for the higher-density/lower-temperature solutions, $\ga$\,1\,$M_\odot$. Although this is not impossible, it remains highly unlikely when considering for example the dust mass estimated for this detached shell, $\approx$3$\times$10$^{-6}$\,$M_\odot$ using dust-scattered light \citep{gonzetal03, olofetal10}, and $\approx$3$\times$10$^{-5}$\,$M_\odot$ using dust thermal emission \citep{schoetal05}. Adopting the 3$\times$10$^{-5}$\,$M_\odot$ dust mass estimate of \citet{schoetal05} of the detached shell (based on thermal emission and therefore more reliable than those estimated from scattered light) and assuming a gas-to-dust mass ratio of 200 (applicable to the circumstellar medium), the detached-shell mass estimate is $\approx$\,6$\times$10$^{-3}$\,$M_\odot$. Even though there are considerable uncertainties in this estimate, it points to the fact that a detached-shell mass above 0.1\,$M_\odot$ is highly unlikely. In this limited range of the ($n_{\rm H_2}$,$T_{\rm k}$)-plane the lower-density/higher-temperature models, only present in the high CO abundance case, are preferred.

The very high CO optical depths obtained for the high-density/low-temperature models are also unlikely since they require $^{12}$CO/$^{13}$CO abundance ratios in excess of 100, while the estimated value for the $^{12}$C/$^{13}$C ratio in the stellar photosphere is 19 \citep{lambetal86}. This value is not very far from the $^{12}$CO/$^{13}$CO line intensity ratios of $\approx$\,10 (see Table~\ref{t:coinputrscl}). Even though strong selective photodissociation of $^{13}$CO cannot be excluded, the most reasonable interpretation is that the $^{12}$CO optical depths are at most moderately high. Assuming LTE conditions, if we take as an indicative criterion that the optical depth of the CO($J$\,=\,\mbox{3-2}) is 1, this will happen at a radial H$_2$ column density of $N_{\rm H_2}$\,$\approx$\,3$\times$10$^{13}$\,$T_{\rm k}^{3/2}$\,$f_{\rm CO}^{-1}$\,cm$^{-2}$. It should be noted that since the $^{13}$CO results speak against high-density/low-temperature models, they also speak in favour of a high CO abundance.

\begin{figure}
\centering
   \includegraphics[width=8cm]{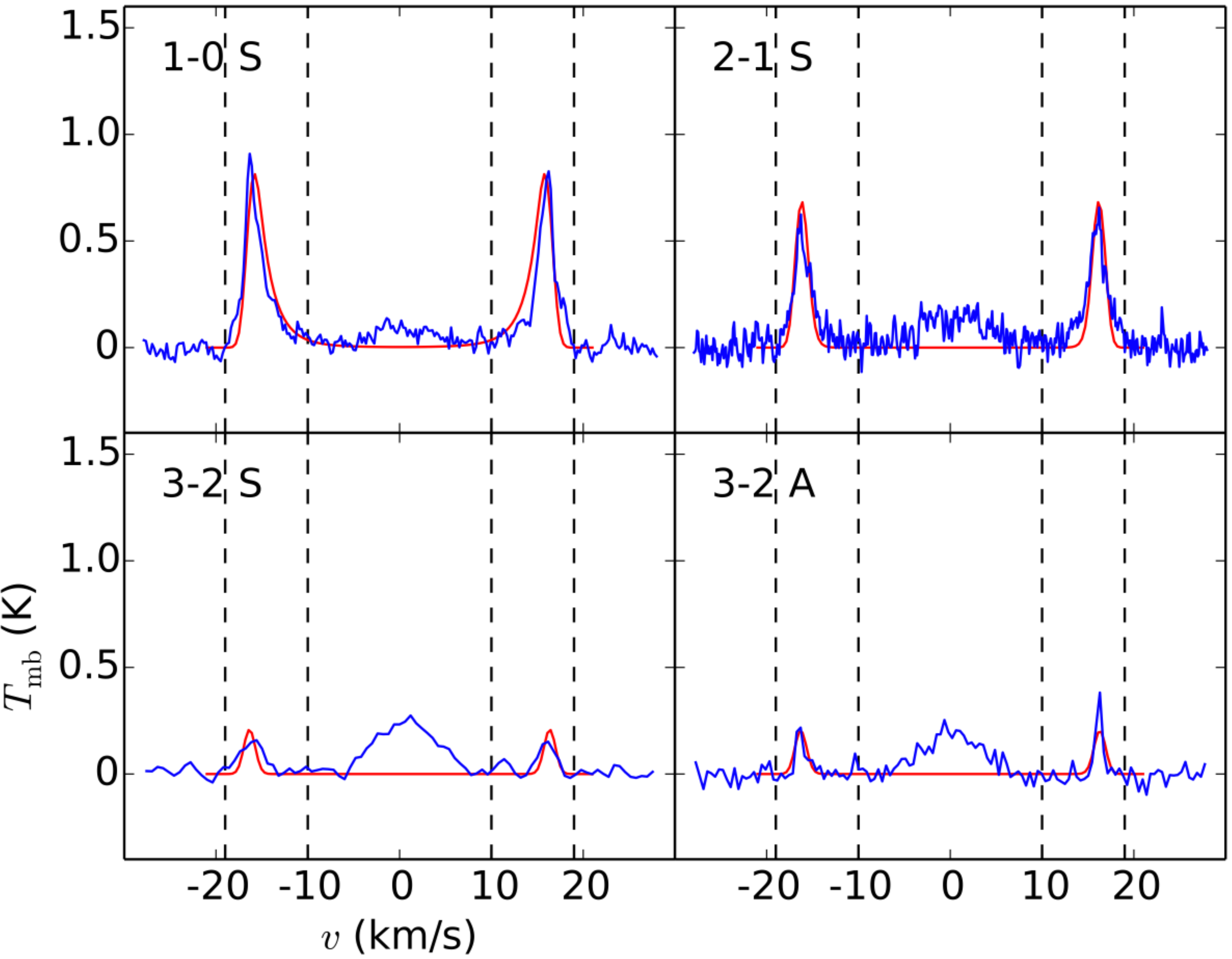}
      \caption{Observed (blue) and model (red) $^{12}$CO lines towards S~Sct (A=APEX, S=SEST). The 2$\times$10$^{2}$\,cm$^3$/100\,K  and $f_{\rm CO}$\,=\,10$^{-3}$ model is used to calculate the model spectra. The velocity scale is centred on the systemic velocity estimated to be 14.9\,km\,s$^{-1}$. The dashed lines indicate the regions used in the fitting to the line intensities. The feature at the systemic velocity originates in the present-day wind and is not part of the modelling.
              }
     \label{f:cospessct}
\end{figure}

\subsection{Best-fit model for a homogenous detached shell}

Using the constraints on detached-shell mass and line optical depths, the models that provide the best fits to the line intensities are the lower-density/higher-temperature ones with a high CO abundance, $\approx$\,10$^{-3}$. The best-fit values are $n_{\rm H_2}$\,=\,(2$\pm$1)$\times$10$^3$\,cm$^{-3}$ (corresponding to $M_{\rm shell}$\,=\,(2$\pm$1)$\times$10$^{-3}$\,$M_\odot$) and $T_{\rm k}$\,=\,100$_{-70}^{+200}$\,K (with the caveat that a temperature above 100\,K is unlikely). These results are in excellent agreement with those of \citet{maeretal15} based on ALMA $^{12}$CO $J$\,=\,\mbox{1--0}, \mbox{2--1}, and \mbox{3--2} data. This corresponds to a mass-loss rate of $\approx$\,10$^{-5}$\,$M_\odot$\,yr$^{-1}$ during the ejection of the detached shell. The best-fit $^{12}$CO/$^{13}$CO and CI/CO abundance ratios are 14$\pm$5 and 0.35$\pm$0.15, respectively. 

However, there are a number of other constraints provided by such processes as photodissociation and photoionisation, and we look at their consequences for our interpretation.

\subsection{Photodissociation of carbon species}
\label{s:dissociation}

Our CI results can, under certain assumptions, further constrain our modelling. Assuming that oxygen is of solar abundance and that all of it is initially locked up in CO, that all other carbon-bearing molecules are dissociated, and that there is no ionisation of carbon, we have for the ratio of neutral carbon and CO in the detached shell,
\begin{equation}
\frac{f_{\rm CI}}{f_{\rm CO}} = \frac{2\, [{\rm O/H}]_\odot \,({\rm C/O - 1}) + f_{\rm d} f_{\rm CO,0}}{(1-f_{\rm d}) f_{\rm CO,0}}
\end{equation}
where $f_{\rm d}$ is the fraction of CO that is photodissociated, and $f_{\rm CO,0}$ is the abundance of CO assuming all oxygen is locked up in CO (i.e. 10$^{-3}$). Adopting the value C/O\,=\,1.34 for R Scl from \citet{lambetal86} we find $f_{\rm CI}$/$f_{\rm CO}$\,=\,0.34 if $f_{\rm d}$\,=\,0 (i.e. $f_{\rm CO}$\,=\,$f_{\rm CO,0}$), which is in excellent agreement with our estimated ratio for the best-fit model. This may be fortuitous, but may, on the other hand, indicate that our assumptions are basically correct. Our results are consistent with the predictions as long as $f_{\rm d}$\,$\la$\,0.1, i.e. only a small fraction of the CO molecules are photodissociated. On the other hand, if carbon ionisation is effective, we get no constraint on the CO abundance in the detached shell (see Sect~\ref{s:ionisation}) and the fact that the CI/CO abundance ratio is 0.35 is fortuitous.

We can make an estimate of the effectiveness of CO photodissociation. Its un-shielded photodissociation rate is 2.6$\times$10$^{-10}$\,s$^{-1}$ \citep{vissetal09}, but -- since it is dissociated in lines -- it is very efficiently self-shielding and in addition H$_2$ lines contribute to the shielding. \citet{vissetal09} have calculated the extra shielding provided by CO itself and H$_2$, and as a guideline CO and H$_2$ column densities of 10$^{16}$\,cm$^{-2}$ and 10$^{19}$\,cm$^{-2}$ (applicable to our best-fit model), respectively, gives an extra shielding by a factor of $\approx$\,25 (although the parameters used in the calculation of this do not match those of our detached shell), meaning a life time of the order of 3000 years, i.e. longer than the age of the R~Scl detached shell. This makes it likely that the CO molecules have survived in the detached shell.

With the assumptions outlined we have shown that the CO in the detached shell of R Scl is affected very little by photodissociation (which is also consistent with the fact that detached CO shells of ages approaching 10$^4$ years, e.g. TT~Cyg and S~Sct, are found, i.e. the CO molecules are still present at these ages). The carbon atoms must therefore come from the full dissociation of the main carbon-bearing species, apart from CO, and whatever carbon that remained in atomic form after the formation of CO and other molecules.  Of course, the propensity for photodissociation is strongly dependent on the clumpiness of the circumstellar medium and this may very well differ significantly from one detached shell to another.

\subsection{Photoionisation of C}
\label{s:ionisation}

The strength of the observed CI line is consistent with a high CO abundance. However, this conclusion requires that the amount of carbon that has been ionised is very limited. 

The observational constraint on the amount of ionised carbon is restricted to a spectrum of the CII line at 158\,$\mu$m in the ISO archive, which is difficult to assess owing to the poor and limited spectral resolution. It corresponds to an integrated line flux of 10$^{-20}$\,W\,cm$^{-2}$, which we will treat as an upper limit. Because of the high energy of the upper level (91\,K), we can expect to see CII line emission only in the higher-temperature models. Using our best-fit model, and assuming conservatively that all the CII line emission comes from the detached shell, this results in an upper limit to the CII/CI abundance ratio of 7 in the detached shell, i.e. not a conclusive upper limit in this context.

We can also make an estimate of the efficiency of carbon photoionisation. The ionisation rate of carbon is estimated to be 3$\times$10$^{-10}$\,s$^{-1}$ (for a normal interstellar radiation field: \citep{drai78}] and \citep{vand88}). With no extinction, the ionisation time scale is only $\approx$\,150\,yr. This means that any un-shielded carbon should have been ionised long before it reached the location of the detached shell. We can make an estimate of the shielding effect. A radial H$_2$ column density of $\approx$\,10$^{21}$\,cm$^{-2}$ through the detached shell corresponds to an extinction $A_{\rm V}$\,$\approx$\,1 \citep{bohletal78} (assuming interstellar conditions). The depth correction is therefore about $e^{3.3}$\,$\approx$\,30 \citep{vand88}, increasing the time scale to $\approx$\,3000 years, i.e. longer than the age of the R~Scl detached shell. However, the best-fit model for a homogenous detached shell has a radial $H_2$ column density of only 10$^{19}$\,cm$^{-2}$, i.e. not enough to effectively protect the carbon against ionisation.

\subsection{The $^{12}$CO/$^{13}$CO ratio in the detached shell and its implications}

There is one more constraint to be discussed, the CO isotopologue ratio. In their study of the $^{12}$CO/$^{13}$CO ratio, based on ALMA images which resolve the emissions, in the circumstellar environment of R Scl, \citet{vlemetal13} found a $^{12}$CO/$^{13}$CO ratio of $\approx$\,60 close to the star and (on average) $\approx$\,20 in the detached shell. They interpreted this in terms of selective dissociation of $^{13}$CO close to the star and a re-formation of $^{13}$CO, through the chemical fractionation reaction, in the detached shell. This requires that the detached-shell temperature is low, especially lower than the difference in ground-state energy between the two isotopologues, which in the case of $^{12}$CO and $^{13}$CO is 35\,K. Thus, these data, if correctly interpreted, advocate a lower-temperature solution.

\subsection{Clumpy medium}

What we have discussed so far cannot be fully reconciled if we assume a homogenous detached shell. In particular, protection against photoionisation of C, and to some extent also the photodissociation of CO, require higher H$_2$ column densities. This may point towards a clumpy medium, where the connection between density, shell width, and mass is offset, and for example higher-density/lower-temperature solutions for a high CO abundance and a limited detached-shell mass become possible. A clumpy medium is also consistent with the small-scale morphology of the detached shells observed in sufficient detail \citep{olofetal00, olofetal10, maeretal12}. 

\citet{bergetal93} and \citet{olofetal96} have presented a model of a clumpy detached shell. The radiative transfer of CO, HCN, CN, CI and CII is solved simultaneously with the physical evolution of a clump as it moves away from the star. The photodissociation of the molecular species and the photoionisation of C are taken into account. \citet{olofetal96} present the results for a 10$^{-5}$\,$M_\odot$ clump of original radius 10$^{13}$\,cm and temperature 750\,K. The evolution is followed over 1.5$\times$10$^4$\,yr. In 2300 years the clump has expanded to a radius of 4$\times$10$^{15}$\,cm, has an H$_2$ density of $\approx$\,2$\times$10$^4$\,cm$^{-3}$, an H$_2$ column density of $\approx$\,1.5$\times$10$^{20}$\,cm$^{-2}$, and a temperature of 35\,K. It was further shown that a thousand such clumps in a detached shell gives a very good fit to both the $^{12}$CO and $^{13}$CO line intensity data towards R~Scl using an abundance ratio $^{12}$CO/$^{13}$CO\,=\,19 (with a much broader shell than the present observational data indicate, but it is expected that this has only marginal effects on the line intensities).

Initially, the CI emission comes from carbon released from HCN (via CN), in this specific case at an abundance 125 times lower than that of CO. Only after $\approx$\,5000 years is there a contribution from photodissociation of CO, i.e. well beyond the age of the R~Scl detached shell. The predicted line intensity ratio between the CI line and the CO($J$\,=\,\mbox{3--2}) line at 2300 years is $\approx$\,0.05 (scaling their result to the CI abundance derived by us and using the APEX beams at the relevant frequencies), remarkably close to the observed line intensity ratio of $\approx$\,0.03 at APEX. Finally, the estimated CII to CI abundance ratio is $\approx$\,0.5, i.e. $\approx$\,30\,\% of the carbon has been photoionised. 

This makes it likely that a clumpy medium will provide an explanation for the observational data, even though we have not made any fine-tuned modelling of such a medium that takes into account for example the much smaller width of the detached shell as shown by the more recent observations.

\section{Conclusions}

The physical properties of detached CO shells around carbon stars are difficult to constrain even when using a substantial number of CO (and $^{13}$CO) lines obtained with single-dish telescopes. It has also proven very difficult to detect any other molecular line emission from these detached shells. Our detection of the CI($^3P_1-\,^3P_0$) line is a step forward and provides additional information. The best fit to the observed CO line intensities, assuming a homogeneous detached shell, is obtained for a detached-shell mass of $\approx$\,0.002\,$M_\odot$ and a temperature of $\approx$\,100\,K if the CO abundance with respect to H$_2$ is 10$^{-3}$ (the $^{13}$CO line emission excludes models with low CO abundances).  The CI($^3P_1-\,^3P_0$) line is detected at an intensity that results in a CI/CO abundance ratio of $\approx$\,0.3 in the best-fit model. This is consistent with the carbon coming from the full photodissociation of all carbon-bearing species except CO, i.e. a high CO abundance is also inferred in this way assuming that the ionisation of carbon is very limited. An ISO spectrum of the CII 158\,$\mu$m line gives an upper limit to the CII/CI abundance ratio of 7 in the case of our best-fit model, i.e. not a conclusive upper limit in this context. It would therefore be worthwhile to redo these observations of the CII line at much higher sensitivity. It is further concluded that a situation that fulfils all the observational constraints, e.g. the required column densities to limit CO photodissociation and C photoionisation, only exists in a clumpy medium. However, the details of such a medium (e.g. the size distribution, density, and temperature of clumps) remain uncertain. Substantial progress can only be made through detailed observations of the small-scale structure of the circumstellar medium using instruments such as ALMA, where the properties of individual clumps are determined.

\begin{acknowledgements}
HO acknowledges financial support from the Swedish Research Council. We are grateful to the anonymous referee for constructive comments, and for pointing out the observations of the CII line at 158\,$\mu$m towards R Scl in the ISO archive.
\end{acknowledgements}



\end{document}